# Combined numerical and experimental estimation of the fracture toughness and failure analysis of single lap shear test for dissimilar welds

Norberto Jiménez Mena[1], Thaneshan Sapanathan[1,*], Pascal J. Jacques[1], Aude Simar[1]

[1]UCLouvain, Institute of Mechanics, Materials and Civil Engineering, IMAP, 1348 Louvain-la-Neuve, Belgium

* Corresponding author e-mail: thaneshan.sapanathan@uclouvain.be

## Abstract

The single lap shear test is widely used to measure the strength of dissimilar welds even though such a test brings limited understanding of the intrinsic weld toughness. The present study proposes a numerical finite element (FE) analysis and experimental characterization of dissimilar joints presenting various microstructures (thickness of the intermetallic layer (IML) and hardness profile). For this purpose, Friction Melt Bonding (FMB) and Friction Stir Welding (FSW) were used to join aluminum AA6061 and Dual Phase steel (DP980). The FE simulations allowed calculating the evolution of the *J*-integral near this notch tip. It shows that crack initiation depends significantly on the plastic properties of the welded metallic alloys around the notch tip and the width of the welded zone, which both are significantly different for FSW and FMB processes. Nevertheless, a similar weld fracture toughness $J_C$ of approximately 1 kJ.m$^{-2}$ is estimated from the analysis for both FMB and FSW. This is three orders of magnitude higher than the fracture toughness of the intermetallic layer, revealing that the plastic dissipation in the Al and steel plates around the crack tip has a major effect on the weld toughness.

## 1. Introduction

Multi-metallic hybrid structures are attractive for engineering applications as they offer unique blends of properties, such as combining high strength and low weight materials in required arrangements [1, 2]. Therefore, manufacturing such metallic hybrid structures has attracted much attention with the associated need of new processing routes (e.g. [3, 4]). The structural performance of such hybrid structures depends on the mechanical properties of the parent materials, the interface properties and the bonding strength of the constituents [5]. Measuring the interface bonding strength between the parent materials is very complex. As a result, characterization of structural properties of hybrid structures is more challenging than classical bulk materials.

A shear strength test based on the ASTM: F1044 [6] standard was used to determine a bond shear strength of hybrid Al/Cu clad rod fabricated using equal channel angular extrusion [7]. However, this standard is mainly applicable to test the joint strength between two flat surfaces when testing an adhesive or cohesive bond between two metallic coated plates [6]. This standard has also been extended for a bond created by any thermomechanical joining process such as sintering or diffusion bonding [6]. However, this standard method is not a reliable measurement of the shear strength of dissimilar welds as a result of a mixed mode of failure.

Nevertheless, to measure Al/steel lap joints strength, single lap shear test remains the most popular test in a lap welding configuration. During a single lap shear test, the joint fails under a mixed mode (modes I and II). Despite the apparent simplicity of the test, the stress fields in the sample are complex due to (i) the misalignment of the tensile loads; (ii) the possible difference in thickness of the dissimilar constituents (Al and steel); (iii) the mismatch in mechanical properties of the dissimilar constituents (Al and steel); (iv) the presence of an intermetallic (IM) at the interface.

Various analytical and numerical models were thus developed to interpret the results of the single lap shear test as reviewed by da Silva *et al*. [8]. Most of the available models focus on adhesive joints with brittle elastic adhesives such as epoxy resins. Since the intermetallic layer (IML) also shows a brittle behavior at room temperature [9], most of these models can be extended to Al/steel welds which present a brittle intermetallic phase at the interface. Particularly, Delale *et al*. developed a model in plane strain conditions that predicts the normal and shear stresses within an elastic adhesive (or IML) between two elastic dissimilar adherents [10]. They predicted that both peeling and shearing stresses concentrate at the extremities of the IML and tend to zero near the weld centerline. The stresses are larger at the extremity corresponding to the side where the higher strength material is located [10].

Moreover, the experimental shear failure loads reported in earlier literature are highly dispersed and difficult to correlate from one case to another [11-23]. Fig.1 compiles the results of several lap shear tests from literature, i.e. the maximum normalized failure load as a function of IML thickness for Al/steel welds with the steel plate thinner (or equal) to 1.5 mm [11-23]. The analysis of the results is usually limited to an evaluation of the maximum load at fracture. This maximum load at fracture is divided by the width of the specimen to identify the bond strength in N.mm$^{-1}$. This simple analysis results from the complex stress fields present in the single lap joint test, i.e. the presence of stress concentration at the extremities of the bonded surface. In Fig.1, the filled and empty marks indicate a failure through the interface (via the IML) and in the base material, respectively. The symbols (circles, squares and stars) correspond to welds without any prior additional layers, with Zn layer at the interface or performed with high Si-containing filler materials, respectively (Fig. 1). It should be noted that the presence of Zn at the interface is either due to previous galvanization or due to addition of a Zn-containing filler material.

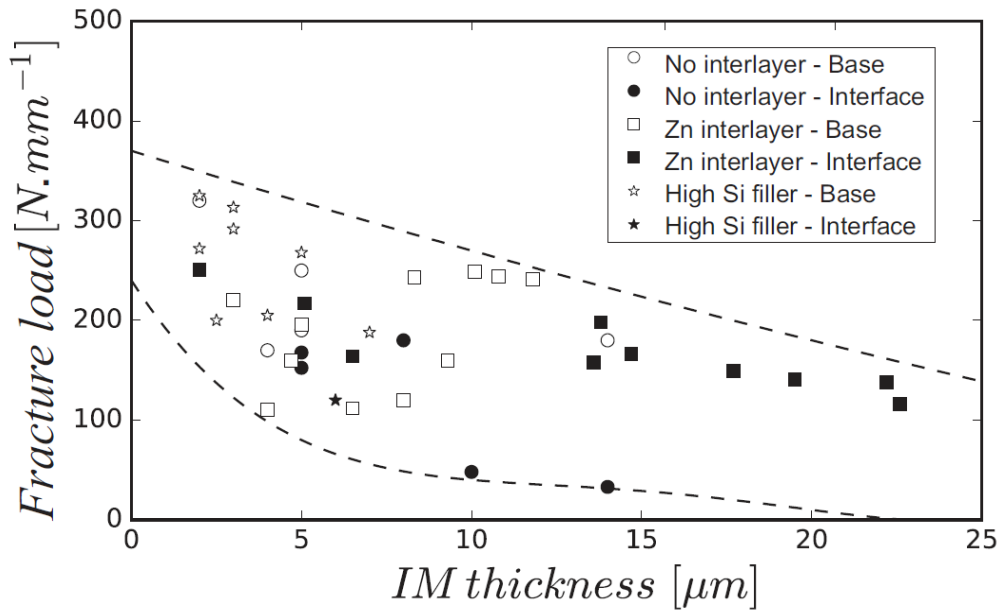

Fig. 1. Normalized maximum failure load observed during the single lap joint test of dissimilar Al/steel welds as a function of the IML thickness. The filled and empty marks denote failure at the interface (via IML) and in the base material, respectively. Circles, squares and stars correspond to welds without any additional layers, with Zn layer at the interface or performed with high Si-containing filler materials, respectively. For comparison purpose, the highest strength values from the literatures of [11-23] are considered. These results include a wide variety of welding processes while having the steel plate thinner (or equal) to 1.5 mm.

Fig.1 shows that the joint strength increases for IML thicknesses below 10 µm. For the same IML thickness, the presence of Zn at the interface does not substantially influence the strength of the weld when compared to a weld without Zn addition. Si additions are known to reduce the IM growth [24], therefore thin IMLs are usually achieved in the presence of Si. Fracture in the base materials are associated to a thin or low strength base material that cannot sustain the load that would be required to cause IML failure.

Since the complex stress distribution in the single lap joint test depends on the sample dimensions, the thickness of the substrates and the welded area among others, a detailed stress analysis is necessary to better understand the load bearing capacity of dissimilar joints under shear loading conditions. Therefore, the present study investigates dissimilar welded lap joints presenting various IML thicknesses and manufactured using Friction Melt Bonding (FMB) or Friction Stir Welding (FSW) processes. Local plasticity and deformation mechanisms during the failure of dissimilar welded joints under shear loading were scrutinized. This combined numerical and experimental investigation provides a better understanding on the failure mechanism of dissimilar lap joints presenting an IML. Furthermore, it allows to estimate a parameter that may be called the toughness of the joint.

## 2. Selection and relevance of the welding processes

Dissimilar joints were processed by two different welding techniques known as Friction Melt Bonding (FMB) and Friction Stir Welding (FSW). FMB is a recently developed welding method, to join dissimilar materials [25, 26]. When manufacturing a FMB Al/steel joint in a lap configuration, the upper steel plate is heated locally by friction. The tool is a flat rotating cylinder (Fig. 2a). The generated heat during the process locally melts the aluminum and facilitates the bond formation with a continuous IML at the interface, with a thickness varying from 5µm to 40 µm [27]. Moreover, experimental observations and numerical simulations of FMB indicate that the temperature during the welding process reaches ~700 °C at the interface [28]. Residual stresses (RS) appear due to the large difference in coefficient of thermal expansion (CTE) between Al and steel [29].

On the other hand, Friction Stir Welding is a widely applied solid-state welding process also adapted to Al/steel dissimilar welds [30]. In an Al/steel lap welding configuration, the tool probe is plunged through the aluminum, while the tip of the probe is also penetrating into the steel (Fig. 2b). Extended description of FSW of dissimilar materials can be found in the review by Simar *et al.* [30]. The observation of IML for dissimilar welding is common for FSW in both butt joints [31-35] and lap joints [36-40]. Some authors have reported that an IML also forms in a dissimilar Al/steel lap friction stir weld without penetration of the probe into the steel plate [41, 42]. It should be noted that the presence of steel fragments into the Al and IMs deteriorates the joint integrity when using large penetration depth of the tool into the steel.

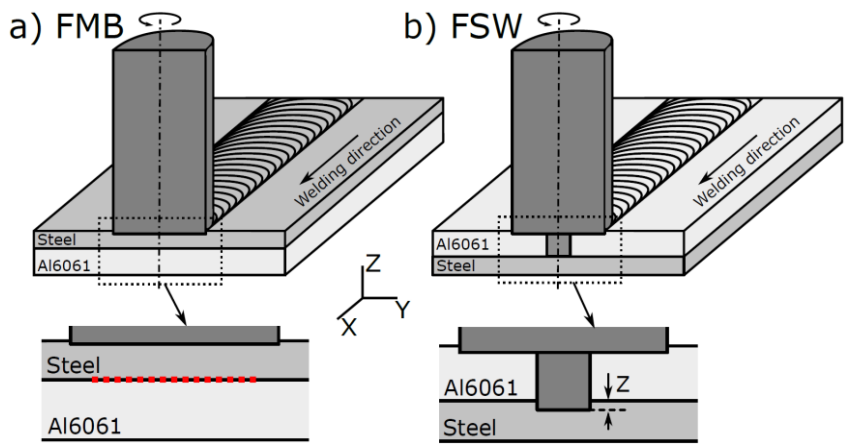

Fig. 2. Schematic representation of the welding processes used in this study (a) FMB and (b) FSW.

Finally, it is also worth noting that the influence of the two welding processes on the IML characteristics have been investigated in earlier studies. For instance, Tanaka *et al.* [31] showed that the tensile strength of dissimilar friction stir butt welds is inversely proportional to the square root of the thickness of the IML. Jimenez-Mena *et al.* performed crack propagation tests on FMB Al/steel joints to measure the toughness of the weld as a function of the IML thickness [43]. It was found that the large toughness measured on specimens with low IML thickness is brought by a larger plastic dissipation in the base materials at the crack tip during propagation. Movahedi *et al.* performed friction stir lap welds between AA5083 and ST-12 mild steel and found that an IML with a thickness lower than 2 µm provide higher joint strength compared to thick IML [12]. Similarly, Lee *et al.* reported that their mechanically strongest welds contain 2 µm of mixed zone along with a 250 nm thick IML [44]. Chen *et al.* reported that the thickness of the IML increases with decreasing welding speed, leading to decreasing joint strength [41]. Both FMB and FSW welding methods were used to establish a clear comparison of the shear strength properties for various IML thicknesses. It is indeed possible since the FSW generates thinner IML compared to FMB, and the thicknesses of IML can be controlled by varying the process parameters.

## 3. Materials and methods

### 3.1 Materials

The base materials, AA6061 and Dual Phase steel have a thickness of 3.1 and 0.8 mm, respectively. Table 1 provides their chemical compositions measured using Agilent® 5100 Synchronous Vertical Dual View (SVDV), inductively coupled plasma - optical emission spectrometry (ICP-OES). The total carbon content ($C^{Total}$) was estimated using Sylab CSbox- infrared analyzer equipped with fusion furnace (IRF 1600F Sylab).

Table 1: Chemical composition (wt.%) of AA6061-T6 and DP980 steel

|  | Al | Fe | Mn | Si | Cu | Cr | Ti | Mg | $C^{Total}$ |
|---|---|---|---|---|---|---|---|---|---|
| AA6061 – T6 | 97.5 | 0.44 | 0.05 | 0.56 | 0.24 | 0.19 | 0.02 | 0.93 |  |
| DP980 steel | 0.02 | 97.45 | 1.94 | 0.21 | 0.01 | 0.19 | 0.02 |  | 0.16 |

### 3.2 Welding conditions

FMB welds were performed with a tool rotation speed of 2000 rpm on a 5 mm thick copper backing plate. The copper backing plate was used here to provide a fast cooling environment that reduces the risk of hot tear formation in the FMB welds [28]. The plunging depth of the tool into the steel plate was set to 0.19 mm. Welding speeds of 100, 150, 200 and 250 mm.min$^{-1}$ were used to generate various IML thicknesses.

For FSW, two different tool pin penetration depths into the steel (i.e. |z| = 0.1 and 0.2 mm, in Fig. 2b) were used to generate various IML thicknesses, as also changed by varying the transverse speed

of the tool. FSW welds were performed on a 5 mm thick stainless-steel backing plate. The rotation speed of the tool was set to 1200 rpm. Welding speeds of the tool were set to 100, 200, 300 and 400 mm.min$^{-1}$.

### 3.3 Mechanical and microstructural characterization

Uniaxial tensile test of DP980 steel and AA6061-T6 base alloys was carried out at a constant crosshead speed of 2 mm.min$^{-1}$, following ASTM E8M-16a standard [45]. The deformation was measured using an extensometer with an initial gauge length of 20 mm. Welded specimens were characterized using the single lap shear test with the dimensions of the specimens provided in Fig. 3a. Specimens were extracted from the welded plate using electric discharge machining (EDM). Three specimens were extracted from each weld. The lap shear tests were performed at a constant crosshead speed of 1 mm.min$^{-1}$. The relative displacement between the grips was used to assess the deformation of the weld. The initial distance between grips was 40 mm. To investigate the crack initiation and propagation, selected welding cases were observed with a high-speed camera, Fastcam SA3 model 120K-M1 with the recording rate of 25,000 frames per second.

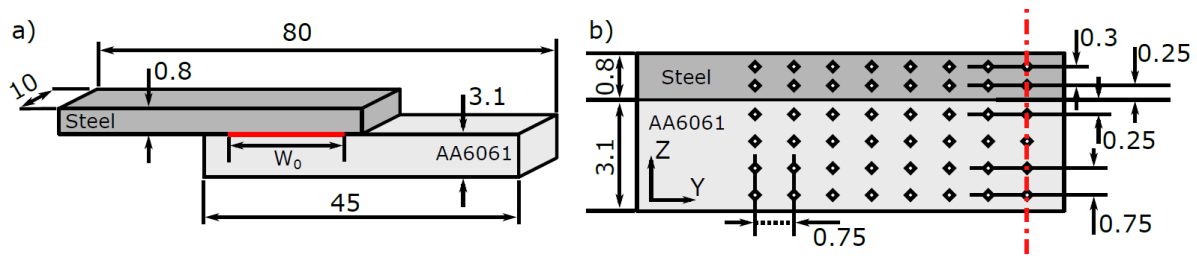

Fig. 3. Schematic illustrations showing (a) geometry and dimensions (in mm) of the samples used for single lap shear test and (b) the grid dimensions (in mm) of the microhardness measurement points on a half-transverse cross section of the welded specimen.

An additional cross-section of the sample was extracted for microstructural observations and microhardness measurements. Microhardness measurements were carried out on transverse sections of the samples to identify the local mechanical properties associated with the change in microstructure during both FMB and FSW processes performed with the transverse welding speed of 200 mm.min$^{-1}$. The hardness measurements were performed using a Vickers indenter, with the weight of 1 kg and 10 s of dwell time according to ASTM E384 – 17 [46]. The spatial distribution of the indents for half transverse cross section of the sample is provided in Fig. 3b. It is worth noting that the heterogeneous properties of steel and aluminum resulted from the thermal history of the process. Therefore, indentation grids of 750 µm × 750 µm and 750 µm × 300 µm were selected for aluminium and for steel, respectively. They capture the property with sufficient accuracy in those materials. This grid size has been selected according to the ASTM E384 - 17 standard that requires 3 times the diagonal indent length between adjacent indents. The thickness of IML was measured from the interface observations by scanning electron microscopy (SEM), operating at an accelerating voltage of 15 kV in both secondary electrons and backscattered electrons modes.

## 4. Numerical modeling

A finite element (FE) model for the shear test was built in Abaqus [47] to understand the development of local stresses and the experimental results from single lap shear tests. Our study focuses on the effect of crack initiation on joint toughness upon accumulated plastic energy in the welded layers. The crack initiates and propagates in the intermetallic layer (IML). In the very hard and brittle IML, there is no evidence of plastic deformation. Plasticity is actually confined in the aluminium plate. An energy release-based argument was thus chosen for crack initiation in the IML. A J-integral calculation is thus used to relate the remote joint toughness via the computed plastic work to the critical energy release rate at the notch tips. A strip yield model would have been more appropriate if there was evidence of plastic flow in the IML. In that case, the currently used elastic IML could have been replaced with a cohesive zone. In addition, this would require a detailed and complex calibration of the traction-separation law of this cohesive zone.

Multiple sets of parameters were used to predict the effect of different dimensions and mechanical properties on the strength of the Al/steel lap joints. Intrinsic model parameters such as (i) the thickness of IML, (ii) the width of the weld seam, (iii) the thickness of the base materials, and (iv) the temper state of the base materials were systematically investigated to identify their effects on the shear strength of the dissimilar joints.

### 4.1 Geometry, mesh and boundary conditions

The geometrical specifications of the FE model are provided in Fig. 4a. Here, $L_0$ and $W_0$ denote the initial distance between clamps during the shear test and the bond length, respectively. $h_{St}$, $h_{Al}$ and $h_{IM}$ correspond to the thicknesses of steel plate, Al plate and IML, respectively. The left and right ends of the IML are represented as notches, which enable to predict the build-up of stresses at the extremities of the IML. The notch tip was placed at the triple junction between the steel, Al and the extremity of the corresponding IML as detailed in Figs. 4a and 4b. An energy criterion (*J*-integral) is then applied for both left and right notches to assess the energy release rate during the crack initiation, i.e., at the fracture load.

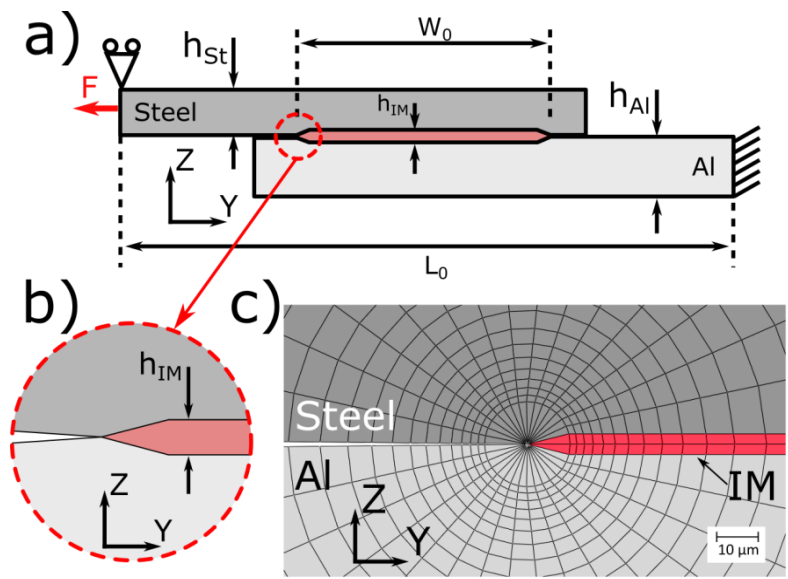

Fig. 4. (a) Geometrical specifications of the single lap joint shear model used in this study, (b) detail of the left notch and (c) mesh around the notch used to calculate the *J*-integral.

The model uses 2D plane strain conditions with an implicit formulation. A structured mesh is implemented around the crack tip using 8 node quadratic quadrilateral elements (CPE8R in Abaqus) in order to calculate the contours of the *J*-integral (Fig. 4c). The first contour is formed by the quarter point elements. Other regions except the crack tip are meshed with 6 node quadratic triangles (CPE6 in Abaqus) to improve the computational efficiency. The end of the Al plate (right side) is fixed in Y and Z directions. On the left end of the steel side, the displacement is constrained in the Z direction. The load is applied on the left end of the steel side parallel to the Y direction (indicated by the load "F" in Fig. 4a).

### 4.2 Constitutive model

The mechanical properties obtained for the base materials are used in the numerical model for the corresponding unaffected zones. Owing to the presence of the heat affected zone (HAZ), the elastoplastic properties of the steel and Al plates are not identical throughout the entire plate. It is thus required to account for the spatial distribution of the local material properties for both steel and Al. A stress-strain curve is assigned to each element accordingly to its hardness value using the correlations described below.

For the Al alloy, the elastoplastic behavior is taken from the work of Jonckheere *et al.* [48] and Lecarme *et al.* [49]. Firstly, the yield strength is assessed for the hardness measurement using the linear relationship [48]:

$$\sigma_y = \frac{(HV - 29.2)}{0.27} \tag{1}$$

where, $\sigma_y$ is the yield strength in MPa and HV is Vickers microhardness. The plastic flow, $\sigma_{pl}$, for each value of yield strength is calculated using a Kocks-Mecking law accounting for the stage III of strain hardening behavior [49]:

$$\sigma_{pl} = \sigma_y + \frac{\Theta_0}{\beta}[1 - \exp(-\beta \varepsilon_{pl})] \tag{2}$$

where $\Theta_0$ and $\beta$ are hardening constants which are fitted from the data available in the work of Hannard *et al.* [50] for a given yield strength.

Similar to Al, the yield strength of the DP steel as a function of microhardness is assessed using a linear relationship obtained from the work of Pavlina *et al*. [51]:

$$\sigma_y = -90.7 + 2.876 HV \tag{3}$$

In the case of steel, Hollomon's law is used to describe the plastic regime of the DP steel. For the sake of simplicity and owing to the lower strains expected in steel plate compared to Al plate, a constant hardening coefficient of n = 0.09, was used for every element. This value was obtained using a mathematical fitting model from the experimental stress-strain curves of the base DP steel. An incremental J2 flow theory with isotropic hardening was used in the FE model to describe plasticity in those materials.

The IML is modelled as a homogeneous linear elastic material without hardness dependency owing to negligible difference in hardness across the IML based on the experimental observations [27]. Based on literature, the elastic modulus and Poisson's ratio of the IML are set as 240 GPa and 0.3, respectively [52].

## 5. Experimental results and discussion

### 5.1 Interface observation

The SEM micrographs given in Fig.5 reveal the IML formation at the interface between the steel and Al sheets for both FMB and FSW processes. It clearly shows that FMB process (Figs. 5a and 5b) leads to thicker IML than FSW (Figs. 5c and 5d). This is due to the higher reactivity of liquid Al in the case of FMB. Indeed, the IML layer growth occurs due to reaction and diffusion of both Al and Fe that depends on the process temperature [27, 53]. The temperatures reached in the case of FMB process is large enough to bring the aluminum to liquid state, while it remains solid in the case of FSW. Moreover, with the increase of the IML thickness, some longitudinal and transverse microcracks appeared in the FMB welds. These microcracks likely formed due to stress relaxation during the cooling of the weld [54].

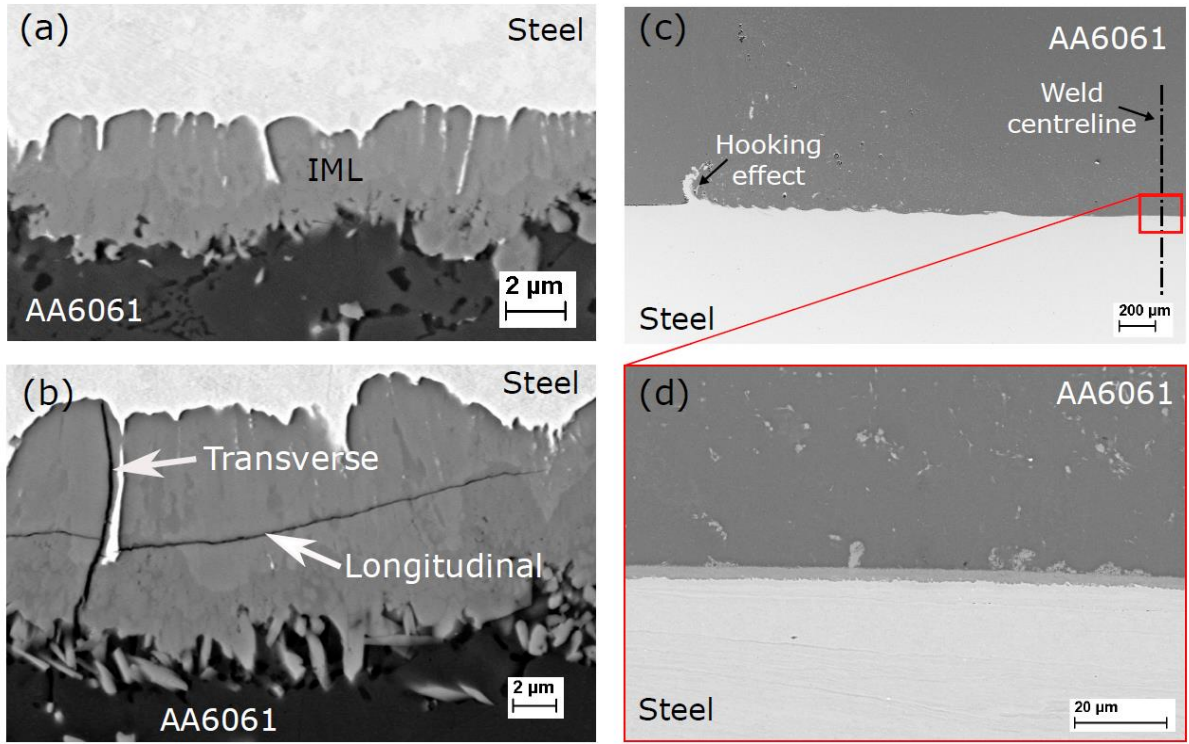

Fig. 5. SEM micrographs of the welded interfaces revealing the IML in the case of FMB processed at (a) 250 mm.min$^{-1}$ and (b) 150 mm.min$^{-1}$ welding speeds. Some longitudinal and transversal microcracks are also present in the case of the FMB weld in (b). (c) and (d) reveal the interface morphology of the FSW welds processed with |z| = 0.2 mm (see Fig. 2b for definition of z) and a welding speed of 300 mm.min$^{-1}$.

Fig. 6 presents the evolution of the IML thickness as a function of the welding speed for three sets of parameters. It is worth noting that the thickness of the IML varies along the interface. For the sake of comparison, the measurements were made at the weld centerline corresponding to the maximum thickness. Fig. 6 shows that the thickness of the IML decreases for an increasing welding speed. For FSW, the IML thickness slightly decreases with the decrease of the tool plunging distance into the steel (*z* in Fig. 2b). It was also identified that the bond length ($W_0$ in Fig. 4a) varied between 10-13 mm and 5-6 mm, for FMB and FSW, respectively. In general, and as expected, the thickness of the IML decreases if the time at high temperature is reduced (i.e. faster welding speed) and the mixing of both materials is limited (i.e. lower tool plunging into the steel for FSW).

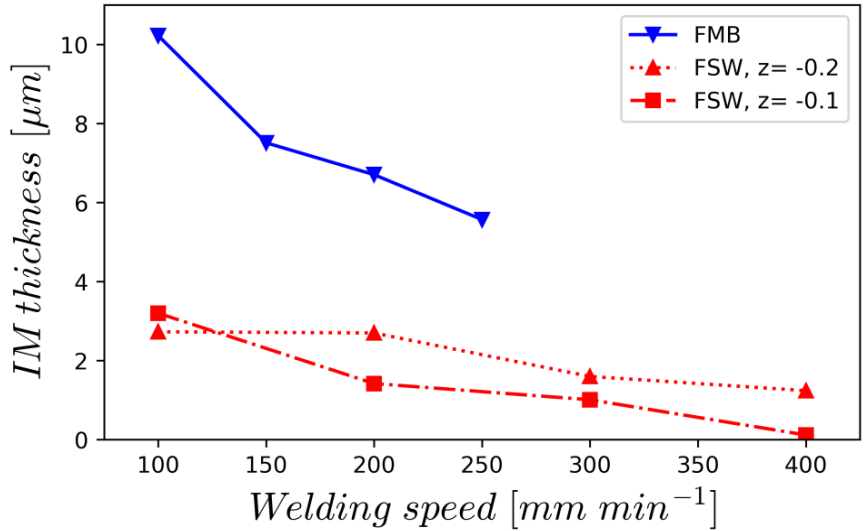

Fig. 6. Evolution of the IML thickness as a function of the welding speed for three sets of parameters.

## 5.2 Mechanical behavior of the joints

The curves of Figs. 7a and 7b show the normalized load (per unit width) as a function of the elongation resulting from the experimental tests and numerical simulations for FMB and FSW specimens, respectively. It is worth noting that common y-axis is used for both Figs. 7a and 7b for comparison purpose. The total elongation [%] was calculated from the change of distance between the grips divided by the initial distance between the grips for the experimental cases. An identical approach for the elongation [%] calculation has been devised for the numerical model. The failure of the specimen occurs suddenly when the load reaches the maximum load.

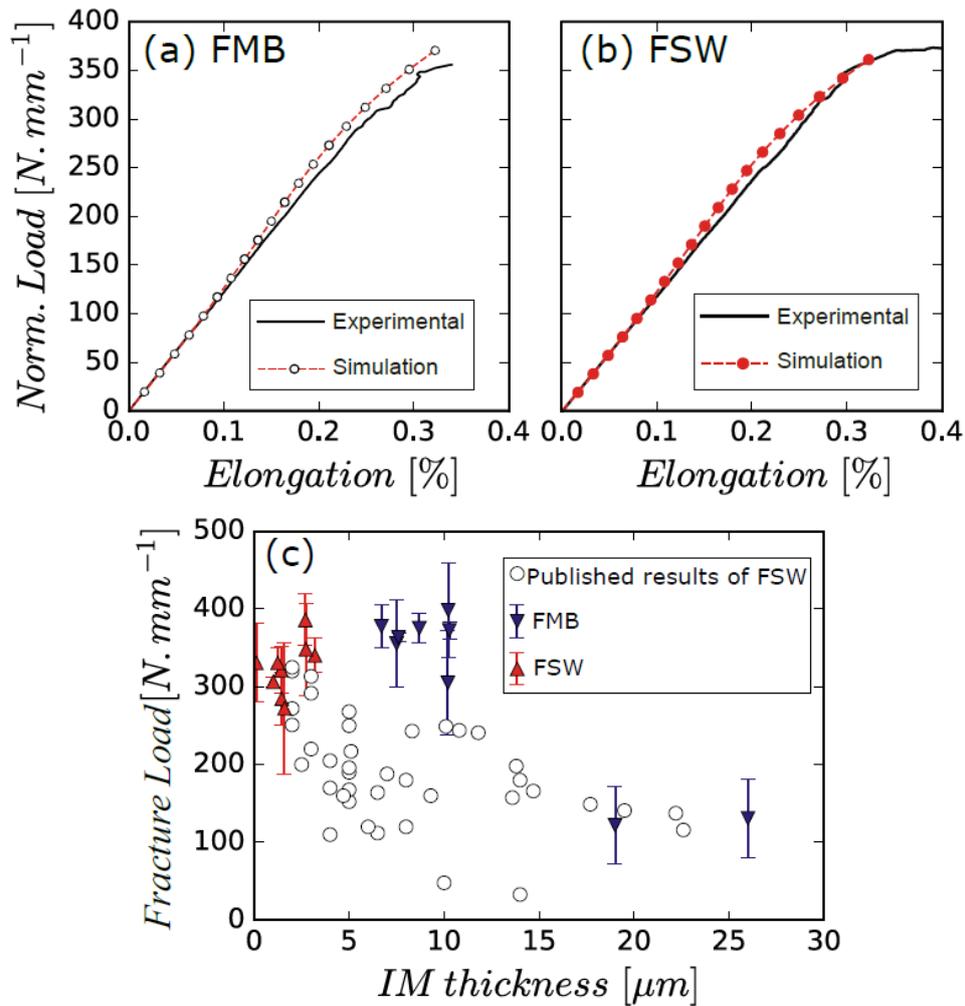

Fig. 7. Normalized load (per unit width) as a function of the elongation obtained from the experimental tests and FE simulations for (a) FMB and (b) FSW [Common y-axis refer to the insets (a & b)]. (c) Fracture load as a function of the IML thickness. The circles represent the results from the literature as summarized in Fig. 1.

Fig. 7c presents the normalized fracture load (fracture load divided by the width of the specimen) for all FMB and FSW joints as a function of IML thickness. The plot also includes data from the literature already provided but with more details in Fig. 1. For the FMB welds, the IML thickness increases significantly (from 5µm to 26µm) with decreasing welding speed. The mechanical tests indicate that the failure load significantly drops to ~125 N.mm$^{-1}$ for samples with an IML thicker than 10 µm. FMB welds with an IML thinner than 10µm exhibit a failure load of 375 ± 15 N.mm$^{-1}$. In addition, there is no significant difference in fracture load with decreasing IML thickness below these 10 µm. A similar behavior is also observed for the FSW samples. The FSW samples present a mean fracture load of 335±45 N.mm$^{-1}$ during single lap shear tests and an IML thickness below 3 µm.

Tanaka *et al.* argued that the toughness of the interface increases with decreasing IML thickness for Al/steel butt welds loaded in tension [31]. They suggested that IML thicknesses below 1 µm are required to achieve a substantial rise in toughness in their butt welds [31]. However, Fig. 7c does not show any significant rise of the fracture load for welds presenting IML thicknesses below 10 µm and loaded in lap shear conditions. In addition, no significant difference is observed between the fracture load of the FSW and FMB welds with IML thickness below 10 µm. A numerical simulation of the lap shear test was thus performed in the present study to understand the local stress development for both FMB and FSW welds, see sections 4 and 6.

### 5.3 Crack propagation

High-speed camera was used to observe the crack propagation of the processed dissimilar welds. Fig. 8 presents snapshots in the case of the FMB specimen welded at 200 mm.min$^{-1}$. It is worth noting that the shear test sample is observed in the same plane as in the schematic of Fig. 4a. With a span time between frames of 40 µs, the crack propagation speed is estimated to be approximately 73 m.s$^{-1}$.

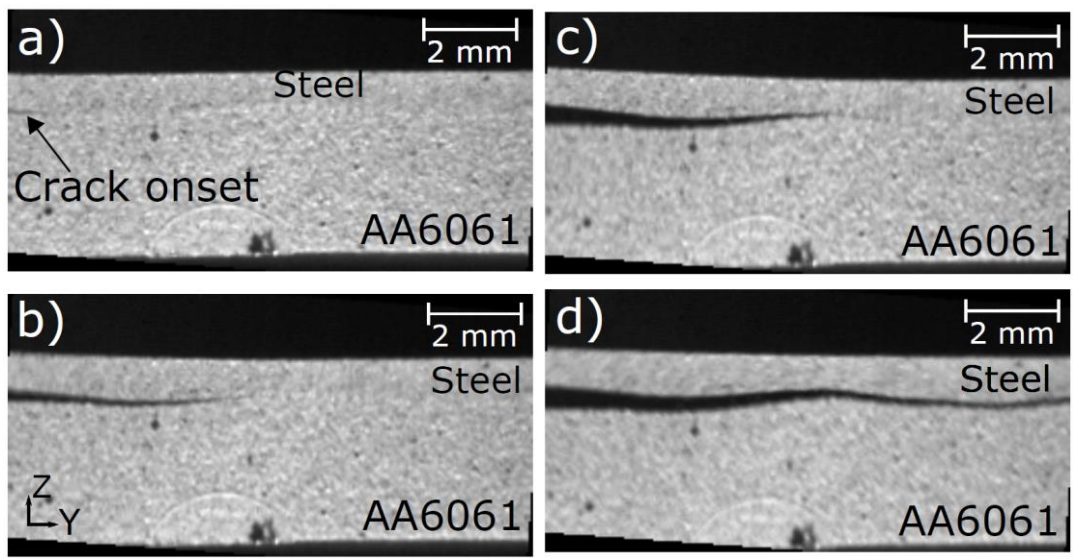

Fig. 8. Frames captured from high-speed camera showing the crack propagation at the interface for an FMB weld performed at 200 mm.min$^{-1}$, at a) 0 µs, b) 40 µs, c) 80 µs, and d) 120 µs. Frame (a) corresponds to the onset of failure. The observed zone is identical to the schematic zoom provided in Fig. 4b.

After the lap shear test, the fracture surfaces were observed for both FMB and FSW samples. The fracture surfaces given in Fig. 9 show that both FMB and FSW samples fractured in a brittle manner as suspected from the sudden load drop in the load-displacement curves (Figs. 7a and 7b). The fracture surfaces of FMB and FSW samples clearly show cleavage facets, which is typical of a brittle fracture that occurred in the IML. The present observations are in agreement with the in-layer crack propagation case described by Akisanya *et al*. [55]. Furthermore, for FMB samples, the fracture surface is flat while it is wavier for FSW samples owing to the deformation caused by the contact of the probe on the steel plate during the welding process.

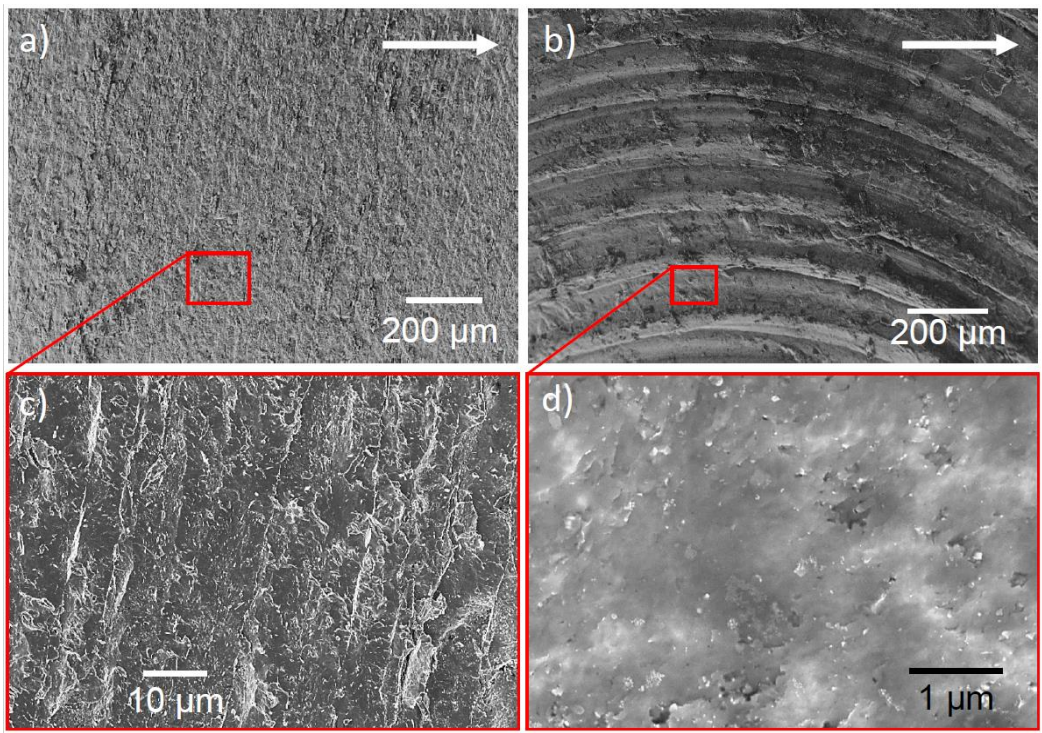

Fig. 9. Fractography observations (in X-Y plane of Fig. 4a) of failure surfaces after single lap shear test for the welds correspond to (a) and (c) FMB specimen welded at 200 mm.min$^{-1}$ and (b) and (d) FSW specimen welded with |z| = 0.1 mm and welding speed of 200 mm.min$^{-1}$. The white arrows in (a) and (b) indicate the direction of crack propagation.

### 5.4 Microhardness measurements

Vickers microhardness values of the base materials are 335±3 and 102±3 HV1 for DP980 steel and AA6061-T6, respectively. The hardness maps of the steel and Al welded zones on the Y-Z cross-section of the weld are provided in Fig. 10 (see Fig. 4a for axis reference). The hardness maps reveal the large differences between FMB and FSW processes attributed to the large difference in thermomechanical cycles. Such differences are particularly notable in the steel plate. For FMB, the steel plate showed an increase in hardness from 335 HV1 to approximately 435 HV1 in the region affected by the process below the tool (Fig. 10a). This observation agrees well with the microstructural changes during the process that has been previously reported [29]. The base steel contains a Dual Phase (ferrite-martensite) structure. During FMB, the steel processed zone reaches a temperature above the austenization temperature of the steel that transforms to either a martensitic or a bainitic microstructure during the cooling behind the tool [29]. The exact steel microstructure formed depends on the cooling rate and thus on the processing conditions. In the case of FSW, the steel region below the tool pin presents a drop of hardness to 260 HV1 (Fig. 10b) due to the low temperature tempering resulting from the heat input.

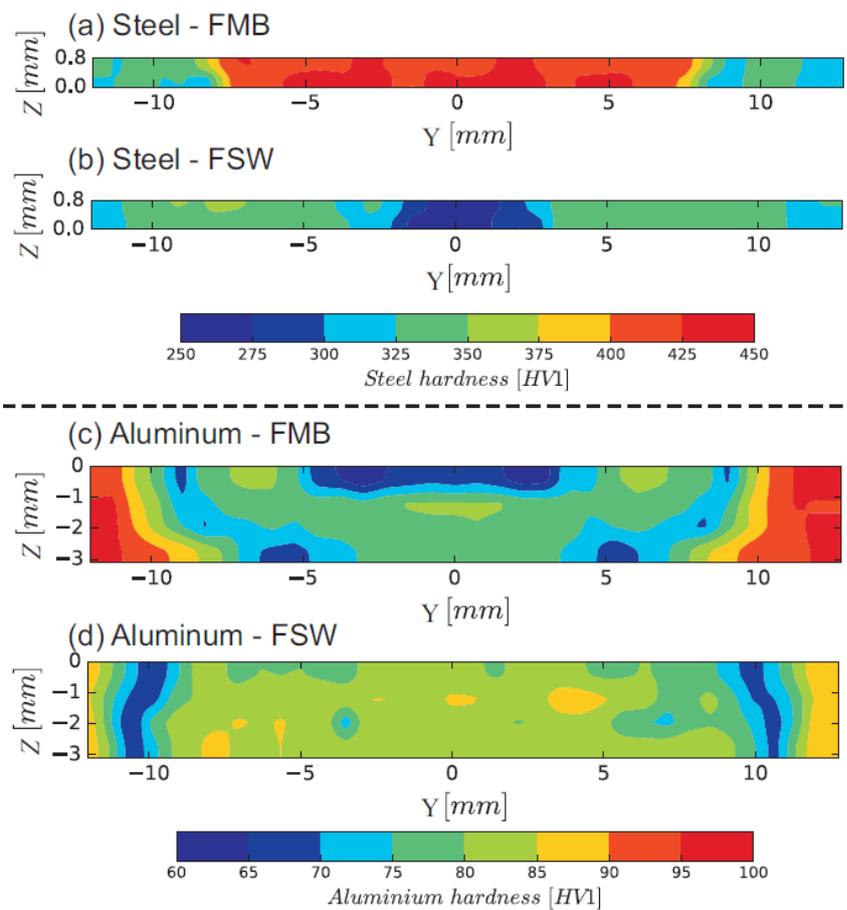

Fig. 10. Micro Vickers hardness (HV1) maps obtained for steel plates in (a & b) and for Al plate (c & d). Maps (a) and (c) correspond to the FMB weld performed at 200 mm.min$^{-1}$, while maps (b) and (d) correspond to the FSW weld performed with |z| = 0.1 mm and welding speed of 200 mm.min$^{-1}$. Note that the legends are different for steel and Al.

On the Al side of the FMB weld, the molten pool shows a drop of hardness to 60-70 HV1 (Fig. 10c) with respect to the base material presenting a hardness of 102 HV1 while the HAZ shows hardness values between 65 and 100 HV1 (Fig. 10c). On the other side, the nugget of the Al plate in FSW shows an average hardness between 80 and 85 HV1 (Fig. 10d). In the HAZ of FSW, the hardness further drops to a minimum of 60-65 HV1 at $y = \pm 10$ mm from the weld centerline.

In the case of AA6061, the microstructure in the T6 state (i.e. the state of the base material) composed of Al grains with $\beta''$ hardening precipitates presents a hardness of 102 HV1. During the FMB process, the aluminum close to the interface melts and resolidifies, causing a drop of hardness to 60-70 HV1 due to the formation of a coarse grain microstructure (see for example in fusion weld, i.e. liquid state welding, seams of AA6061 performed by Peng *et al*. [56]). In the HAZ of Al in FMB welds (i.e. adjacent to the molten pool), the drop of hardness can be attributed to the dissolution-precipitation process of the $\beta''$ precipitates [57]. At an increasing distance from the weld centerline, the decrease of hardness is due to the coarsening or partial dissolution of $\beta''$ precipitates, their transformation into $\beta'$ and $\beta$ precipitates and the coarsening of the latter. In the case of FSW, the hardness profile presented in Fig. 10d is similar to Jonckheere *et al.* [48] who performed AA6061/AA6061 friction stir welds. The temperature reached in the nugget during the welding process was sufficiently high to dissolve the hardening precipitates and to form new GP zones. Similar arguments to FMB joints can be used to explain the drop of hardness in the HAZ of FSW joints (Fig. 10d), reaching a minimum at $y = \pm10$ mm from the weld centerline.

## 6. Numerical predictions

As mentioned earlier, despite the simplicity to carry out single lap joint tests, the analysis of the results is somewhat challenging owing to the complex stress fields developed in the specimen followed by sudden fracture. Therefore, a numerical framework is established to understand the crack initiation and to perform a reliable benchmark for the single lap joint test for FMB and FSW joints. The load at fracture, the dimensions of the IML and the hardness maps are used as inputs for the FE model. The *J*-integral is evaluated at the notch region of the single lap-joint specimen (Figs. 4b and 4c).

### 6.1 Finite element simulation

According to the interface and IML observations (Figs. 5 and 6), mean values of $W_0$ and $h_{IM}$ obtained from various welding conditions for both FMB and FSW are used in the FE model along with the mean fracture loads provided in Fig. 7c. These values are listed in Table 2. It is worth noting that the geometrical parameters listed here are identified based on the sensitivity analysis of IML thickness ($h_{IM}$) and weld width ($W_0$) (see section 6.3 for more details).

Table 2: Load and dimensions used for the FE model for both FMB and FSW. See Fig. 4 for the definition of the dimensions.

| Variables | FMB | FSW |
| --- | --- | --- |
| $h_{IM}$ [µm] | 7.5 | 2 |
| $W_0$ [mm] | 11.5 | 5.5 |
| Fracture load [N.mm$^{-1}$] | 375 ± 15 | 335 ± 45 |

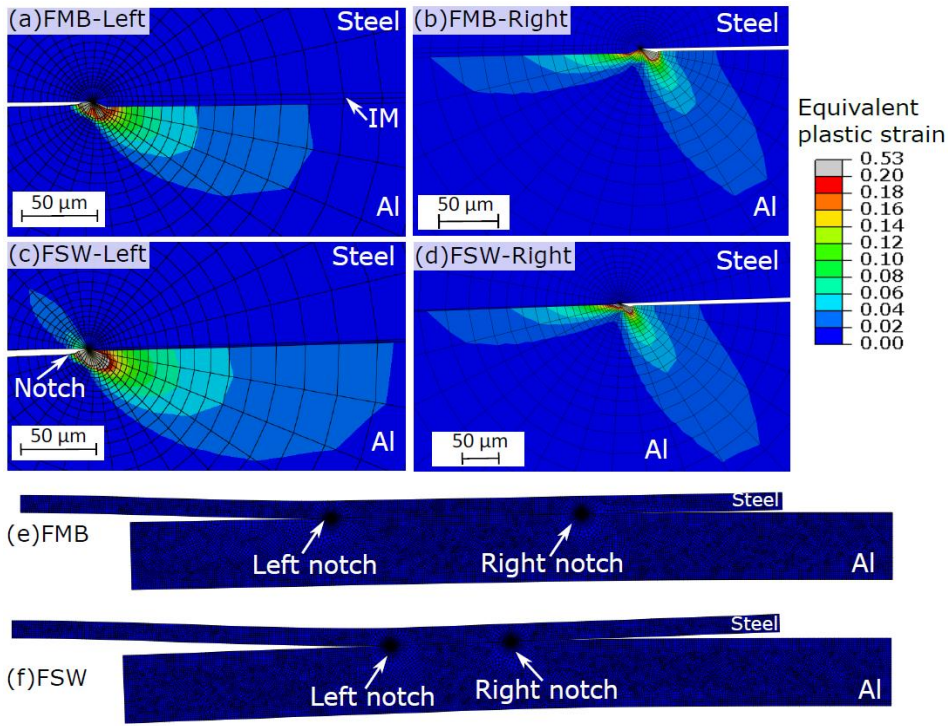

Fig. 11. Equivalent plastic strain distribution during the lap shear test at an applied load corresponding to fracture (Table 2) for (a) FMB-Left notch (b) FMB-Right notch (a) FSW-Left notch and (d) FSW-Right notch. Deformed shape of the (e) FMB and (f) FSW joints at fracture load.

Figs. 11a and 11c show the predicted equivalent plastic strain distribution at fracture load around the most critical left notch (for reference to position, see Figs. 11e and 11f) obtained from the numerical simulation of lap shear test for FMB and FSW joints, respectively. Figs. 11b and 11d illustrate the equivalent plastic strain around the right notch for FMB and FSW joints, respectively. The deformed shape of both FMB and FSW are given in Figs. 11e and 11f, respectively. The predicted plastic strains around the notch regions are high compared to other regions of the test sample. The steel plate in FSW exhibits negligible local plasticity around the notch region (Fig. 11c), while no plasticity is observed in the steel plate for FMB samples (Fig. 11a and 11b). The notches present crack blunting in the Al plate and this blunting is slightly larger for FSW compared to FMB.

### 6.2 *J*-Integral

The *J*-integrals are evaluated at both notches (left and right) for FMB and FSW processes. Although, on a theoretical point of view, using classical $J2$ plasticity for a heterogeneous problem could violate path-independency in the plastic zone, one can see that the plasticity around the notch is limited to a zone of typically 200 µm size and most plasticity is mainly confined in the aluminium plate (Fig 11). Besides, the material properties used in the model were obtained from hardness measurements, and for the aluminium the hardness measurements were obtained using a grid size of 750 µm (section 5.4). Thus, the material properties in the model near the plastic zone (i.e. 750 µm around the notch) are actually constant. Owing to these practical details, *J* integrals can be safely computed around the notches. It is also worth noting that the elastic properties along the direction of the crack is constant (intermetallic is modelled as a homogeneous elastic material). See for e.g. in [58] for more detail about a similar analysis for quasi-static fracture problem in functionally graded materials.

The *J*-integrals calculations are depicted in Fig. 12 as a function of the contour distance from the notches. At a given load, the value of the *J*-integral quickly increases with the distance from the notch and then becomes stable (Figs. 12a and 12b). Figs. 12c and 12d show the evolution of the evolution of steady state *J* values as a function of the load on the specimen for FMB and FSW, respectively. These J-integral evaluations given in Figs. 12c and 12d were extracted at 300 µm, which is beyond the plastic zone (corresponding to the constant plateau) and located solely in the linear elastic material regimes, in Figs. 12a and 12b. During the test, the *J* value at the left notch is approximately twice the *J* value at the right notch (Figs. 12c and 12d). The larger results of the *J*-integral for the left notch agrees with the high-speed camera observations of failure where the crack initiates on the left notch and is observed to propagate towards the right (Fig. 8). It is also worth noting that *J* increases at a higher rate with respect to the load for FSW than for FMB. The plastic zones are also larger in FSW (244 µm and 270 µm for left and right notches, respectively) than in FMB (194 µm and 179 µm for left and right notches, respectively) welds at their respective fracture loads (marked in Figs. 12a and 12b).

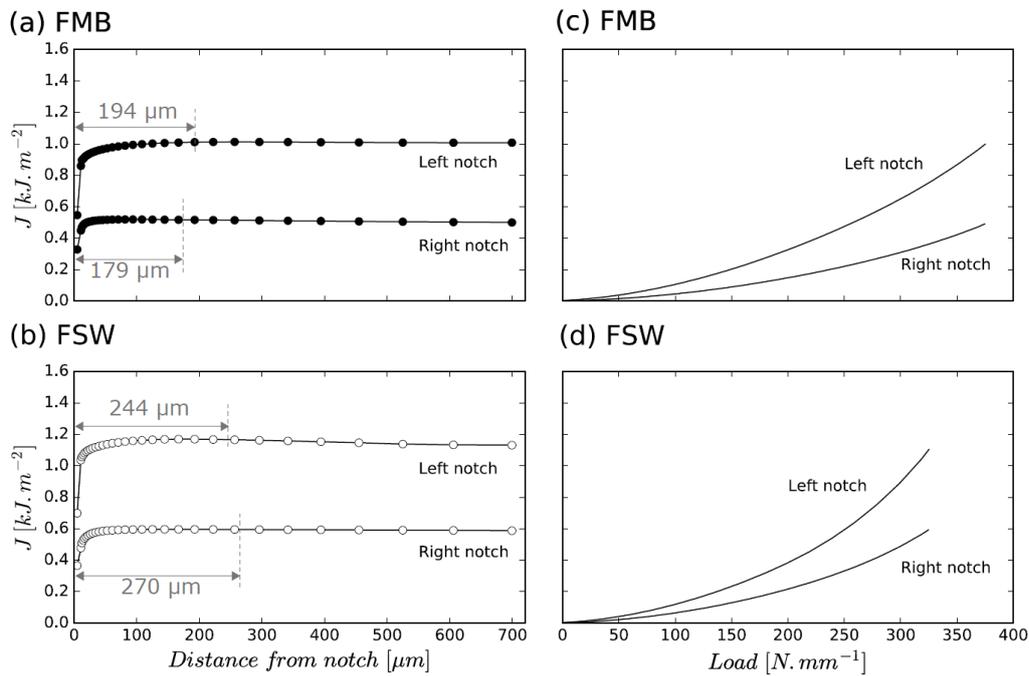

Fig. 12. Evolution of the *J*-integral at the fracture load as a function of the distance to the notch for (a) FMB and (b) FSW. Evolution of the *J*-integral as a function of the load for (c) FMB and (d) FSW. [Common x-axis refers to the insets (a & b) and (c & d), while common y-axis refers to the insets (a & c) and (b & d)]. The plastic zone sizes are also marked in a and b. J-integral evaluations given in c and d were extracted at 300 µm which is beyond the plastic zone (i.e. on the constant plateau).

**6.3 Sensitivity analysis of IML thickness and weld width on *J***

The influence of the IML thickness ($h_{IM}$) and weld width ($W_0$) (see Fig. 4 for reference to geometrical features) on the computation of *J* integral are investigated to decouple these effects and thus understand the differences between FSW and FMB processes. Indeed, FMB joints present a thicker IML and a wider bonded width than FSW (Table2). Thus, two different IML thicknesses, i.e. 2 µm (thin) and 20 µm (thick), were simulated while keeping identical other parameters used for FMB welds. The results revealed a difference lower than 0.1% in the value of *J* between these two cases and thus present no interest to be represented graphically. In conclusion, a thickness of the IML between 2 µm and 20 µm has a negligible influence on the evolution of the *J*-integral.

Three lap shear tests for FMB welds presenting different welded widths ($W_0$) (see Fig. 4 for reference to geometrical features) were also simulated to investigate the sensitivity to this parameter. Fig. 13 shows the evolution of the *J*-integral around the notch tip as a function of the load on the left notch for FMB welds with different weld lengths ($W_0$ = 10, 11.5 and 13 mm, respectively). These J-integral evaluations were extracted at 300 µm which is beyond the plastic zone. For this particular simulation, the heterogeneous material properties depending on the hardness maps for FMB (Fig. 10) were obtained using mathematical model, as a function of position for $W_0 = 13\ mm$ and $W_0 = 10\ mm$, compared to the calculations in Fig. 12 where $W_0 = 11.5\ mm$ (red curve in Fig. 13). The simulations reveal that the *J*-integral around the notch tip increases at a slightly higher rate when decreasing the values of $W_0$.

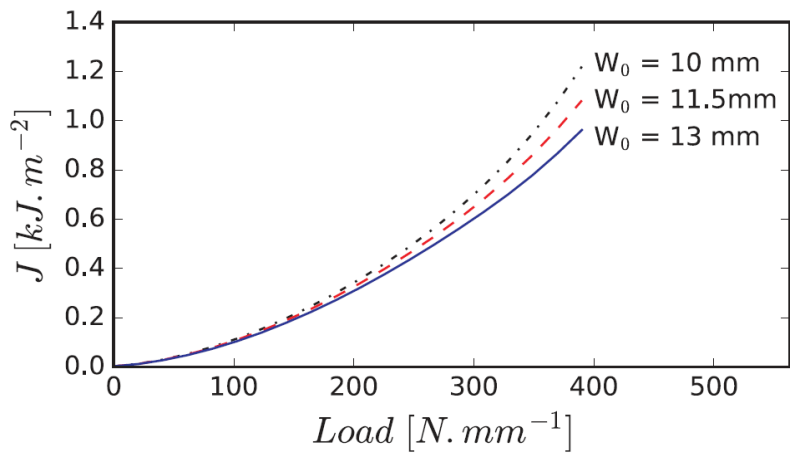

Fig. 13. Effect of weld width ($W_0$) on the evolution of the *J*-integral (on the left notch) as a function of the applied load during the lap shear test of FMB samples. J-integral evaluations were extracted at 300 µm which is beyond the plastic zone (i.e. on the constant plateau).

These results show that part of the reason why FSW welds present a faster rise of the *J*-integral around the left notch tip when compared to FMB, can be associated to the smaller $W_0$ value of these welds (Table 2). However, the hardness distribution (Fig. 10) also contributes to the redistribution of plasticity near the notch tip, which also causes a change in this *J*-integral evolution. Indeed, the distribution of plasticity is clearly different in both processes (Fig. 11).

### 6.4 Estimation of the toughness of the weld

The sudden failure of the single lap joint specimens indicates that the main phenomenon dominating the failure of the Al/steel welds is the initiation of a crack in the IML. Simulations presented in sections 6.1 - 6.3 suggested that large stresses build up in both steel and aluminum at the left notch during loading. When a critical stress is reached, the crack initiates and quickly propagates by releasing the stored elastic energy. Since the crack propagates from the left to the right, the *J*-integral values around the left notch, where the crack initiates, are used as fracture criteria. The *J*-integral values around the left notch are 1.0 and 1.1 kJ.m$^{-2}$ for FMB and FSW (Figs. 12c and 12d), respectively, for the corresponding fracture loads (375 N.mm$^{-1}$ for FMB and 335 N.mm$^{-1}$ for FSW).

According to the work of Akinsaya *et al*., in-layer crack propagation could be used to measure macroscopic toughness, without considering the type of loading and the consequence of mode mixity [55]. The macroscopic fracture toughness given by $J_C$ and calculated *J* values in Figs. 12c and 12d are thus comparable for both FMB and FSW, regardless of the mode mixity during failure. In addition, the simulation presented in sections 6.1 - 6.3 have shown a negligible influence of the IML thickness on the *J* value. It can be concluded that a $J_C$ of about 1 kJ.m$^{-2}$ is representative of a wide variety of dissimilar welds for the two selected base materials.

The idealized situation of the model can of course cause inaccuracies in the evaluation of the weld fracture toughness. In this analysis, the surface was considered as perfectly flat, which is not the case in the actual welds. In particular, the hooking effect that appears at the FSW interface (see Fig. 5c) can affect the stress-strain field at the notch. While loaded in shear, the presence of some surface roughness can bring part of the IML into a compressive stress state that can postpone the initiation of the crack. Nevertheless, the present model allows a useful evaluation of the order of magnitude of weld toughness.

### 6.5 Contribution of IM and substrate properties on the fracture toughness of weld

The assessment of $J_C$ accounts for the strain accommodation in the Al plate, steel plate and IML. Assuming linear elastic properties and plane strain condition, the $J_C^{IM}$ of the IM phase can be estimated using Eq. (4) [59]:

$$J_C^{IM} = \frac{K_{IC}^2}{E'} \tag{4}$$

where $K_{IC}$ and $E' = E/(1 - v^2)$ are the critical stress intensity factor and the modified Young's modulus of the IM phase to account for the plane strain condition. $v$ is the Poisson's ratio. Using $K_{IC} = 0.51$ MPa.m$^{1/2}$ [60], $E = 240$ GPa [52] and $v = 0.3$, Eq. (1) yields $J_C^{IM} \approx 1$ J.m$^{-2}$. This value is three orders of magnitude lower than the predicted $J_C$ from the FE simulations. Hence, the contribution of the straining of the steel and Al plates is dominant over the contribution of the IML. The predicted $J_C$ from numerical simulation thus provides an indication on the amount of energy that the Al and steel plates dissipate prior to the crack initiation.

As discussed by Simha *et al.*, in elastoplastic bodies, the *J*-integral near the notch tip, $J_{tip}$, should be differentiated from the *J*-integral that contains the whole plastic zone around the notch, $J_{far}$ [61]. In the study of crack propagation, $J_{tip}$ tends to be equal to the intrinsic toughness of the IML, while $J_{far}$ can be interpreted as a global driving force and should be used to characterize the toughness of the specimen. This is in agreement with the *J* curves as a function of distance from the notch provided in Figs. 12a and 12b. Close to the notch tip, *J* is lower due to the presence and significant influence of IML, thus providing lower $J_{tip}$ value compared to $J_{far}$. As the distance from the notch tip increases, *J* converges to $J_{far}$.

## 7. Conclusions

Experimental and numerical finite elements (FE) investigations of lap shear test have been carried out for dissimilar Al/steel welds to understand the complex failure mechanism under shear loading. Friction

Melt Bonding (FMB) and Friction Stir Welding (FSW) processes are used to produce welds with various thicknesses of intermetallic layer (IML). Based on this study the following conclusions can be drawn:

(i) The presence of IML determines the mode of failure of the specimens tested in single lap shear test. The failure occurs by the propagation of a crack through this IML.

(ii) The maximum load measured in single lap joint test does not increase with the decreasing IML thickness below 10 µm. FMB presents a failure load of 375±15 N.mm while having an IML thickness between 6 and 10 µm. FSW welds present a failure load of 335±45 N.mm$^{-1}$ while the IML thickness remains below 3 µm;

(iii) FE simulations and high-speed camera observations reveal that the failure of the specimen is dominated by the crack initiation while the propagation is extremely fast once the crack has initiated. The presence of a sharp notch in lap welds favors a built up of stresses in this location and thus the crack initiates there.

(iv) The *J*-integral evolution with applied load is different for FSW and FMB welds due to the difference in local plastic properties distribution and width of the welded zone in these two welds. The simulation has indeed revealed that *J*-integral evolution with applied load depends slightly on the width of the welded zone. In contrast, the FE simulations have revealed that the intermetallic thickness between 2 and 20 µm has negligible influence on the evolution of the *J*-integral.

(v) A crack initiates in FMB and FSW when the remote *J*-integral reaches a critical value $J_C$ of approximately 1 kJ.m$^{-2}$ for both FMB and FSW. Although, FMB and FSW welds exhibit significant differences in hardness distribution, width of joint (and thus of *J*-integral evolution) and IML thickness, they still lead to similar weld fracture toughness.

(vi) The fracture toughness of the weld (defined as the $J_C$ extracted from the FE simulations) is three orders of magnitude higher than the fracture toughness of the intermetallic. This highlights the importance of the plastic dissipation in the Al and steel plates around the crack tip.

**Acknowledgements**

The authors acknowledge the WALInnov ALFEWELD project (convention n°1710162) funded by the Service public de Wallonie Economie Emploi Recherche (SPW-EER). NJM acknowledges the financial support of FRIA, Belgium. NJM and TS also thank Prof. Thomas Pardoen (Institute of Mechanics, Materials and Civil Engineering, UCLouvain), Dr. Frédéric Lani (previously at UCLouvain) and Dr. Frederik Van Loock (Institute of Mechanics, Materials and Civil Engineering, UCLouvain) for fruitful discussions on J2 plasticity calculations and interface modelling for dissimilar metallic/intermetallic joints. TS acknowledges the financial support of National Fund for Scientific Research (FNRS), Belgium. AS acknowledges the financial support of the European Research Council for a starting grant under grant agreement 716678, ALUFIX project.

**Data availability**

The raw/processed data required to reproduce these findings cannot be shared at this time as the data also forms part of an ongoing study.